# Concept for a CMOS Image Sensor Suited for Analog Image Pre-Processing

Lan Shi,
Christopher Soell,
Andreas Baenisch,
and Robert Weigel
Institute for Electronics Engineering
University of Erlangen-Nuremberg
Cauerstr. 9, 91058, Erlangen
Germany
Email: {lan.shi, christopher.soell,
andreas.baenisch, robert.weigel}
@fau.de

Jürgen Seiler
Chair of Multimedia Communications
and Signal Processing
University of Erlangen-Nuremberg
Cauerstr. 7, 91058 Erlangen
Germany
Email: seiler@lnt.de

Thomas Ussmueller
Institute for Mechatronics
University of Innsbruck
Technikerstrasse, 6020 Innsbruck
Austria
Email: thomas.ussmueller@uibk.ac.at

*Abstract*—A concept for a novel CMOS image sensor suited for analog image pre-processing is presented in this paper. As an example, an image restoration algorithm for reducing image noise is applied as image pre-processing in the analog domain. To supply low-latency data input for analog image pre-processing, the proposed concept for a CMOS image sensor offers a new sensor signal acquisition method in 2D. In comparison to image pre-processing in the digital domain, the proposed analog image pre-processing promises an improved image quality. Furthermore, the image noise at the stage of analog sensor signal acquisition can be used to select the most effective restoration algorithm applied to the analog circuit due to image processing prior to the A/D converter.

## I. INTRODUCTION

In the current digital camera market, imaging systems are required to have high performance but low energy consumption, at high speeds but small size. A digital camera with a CMOS image sensor offers significant advantages over traditional CCD in terms of low-power consumption, low-voltage operation use and monolithic integration [1].

A computational CMOS image sensor integrates imaging elements and image processing circuitry at the focal plane. Njuguna and Gruev have divided the state-of-the-art CMOS image sensors into two groups [2]. The first group incorporates imaging and digital signal image processing. The sensor image signal firstly is digitized, then processed with image processing algorithms in the digital domain on the same chip. The second group incorporates imaging and mixed-signal image processing. The digital programmable analog computational circuitry is incorporated at the pixel level or as part of the A/D converter. Our target is the development of a novel CMOS image sensor that incorporates imaging and analog image pre-processing after analog signal acquisition and before A/D conversion.

More and more electronic circuits and functions are being designed economically in the digital domain. There are many automation techniques at various levels of circuit design. The module's application is flexible, and can be functionally integrated to a single chip on a large scale. However, an analog implementation can be superior to a digital approach in terms of speed, area and power consumption for some applications [3]. For example, an analog charge-domain FFT design for a software-defined radio receiver front-end in 65 nm CMOS operates at speeds $5\times$ faster than the state-of-the-art digital design, and consumes $130\times$ less energy [4]. An analog CMOS circuit design could thus greatly increase the hardware efficiency.

Moreover, the performance of CMOS analog circuit technology has also advanced to the point that most discrete analog components, e.g., in the discrete time domain, can be produced using current CMOS technology. Due to the fact that a metal-oxide semiconductor field effect transistor (MOSFET) can be used in chip circuits for analog switches that have very high off-state impedance, this can result in excellent sample-and-hold and switched capacitor circuits, which are very important building blocks for CMOS image sensors [5]. Therefore, the challenge to quickly improve the image sensor signal quality on the CMOS circuit in the analog domain remains.

To apply a 2D-filter to restore image signals in the spatial domain [6] on the analog circuit, a signal acquisition method for obtaining neighboring information in parallel is required. In a study by Duois et al. [7], a CMOS sensor with a massively parallel architecture was designed, and a Sobel and a Laplacian filter were implemented on the circuit. The speed attained was up to 10,000 frames per second. However, the sensor's fill factor only reached 25% because of the additional complex computation circuit in each pixel.

Our proposed concept of a novel image sensor provides a new image signal acquisition method and supports a low delay analog image pre-processing prior to the A/D Converter. It has the potential to maintain a high sensor fill factor, because the

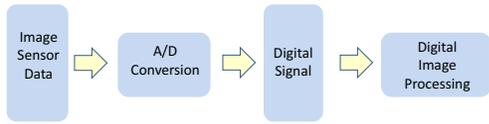

(a) Current image sensor signal processing architecture

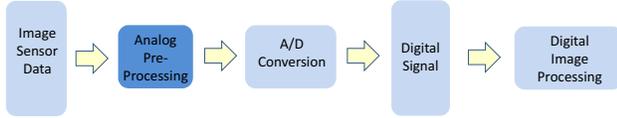

(b) Proposed image sensor signal processing architecture

Fig. 1. Image sensor signal processing work-flow

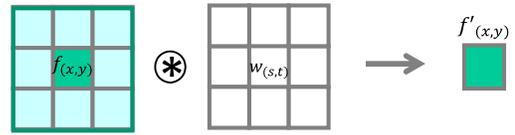

(a) Conceptional filter with $3 \times 3$ pixels

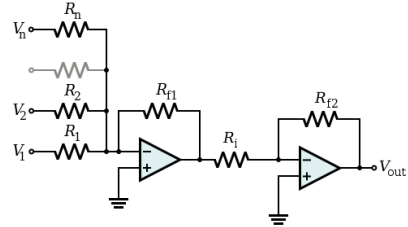

(b) Circuit realisation with operation amplifiers

Fig. 2. Linear spatial filters

sensor pixel does not include any computing circuitry. Instead of a conventional, line-by-line sequential readout scheme, the proposed readout is two-dimensional and suitable for linear spatial filtering algorithms.

With the proposed concept, two spatial filters of small size, such as an average filter and a binomial filter with a size of $3 \times 3$ pixels, are tested as analog pre-processing algorithms to reduce random image noise. The output image, which is firstly processed in the analog domain and then digitized, has a better image quality than that which is firstly digitized and then processed in the digital domain. Furthermore, which filter is efficient on which analog circuit is determined by the random image noise after analog data acquisition at the front end of an image sensor. Analog 2D signal processing prior to the A/D converter allows a better custom-tailored and more efficient filtering orientated to the detected kind of determined noise.

This paper is organized as follows: Section II presents the employment of analog image pre-processing. Section III shows the 2D data acquisition method developed for low delay analog image pre-processing. Section IV compares the image pre-processing in the analog and digital domains and discusses the results of the output image quality. Section V compares two spatial filters as analog image pre-processing for reducing image noise and discusses the effective range of both filters to support the analog circuit design in future work. The paper concludes in Section VI and discusses prospective further work.

## II. ANALOG DOMAIN IMAGE PRE-PROCESSING

The pipeline of conventional image signal processing is shown in Fig.1(a). The image sensor data is first an analog electrical signal (charge, voltage or current). After the data acquisition of the image sensor array, this analog signal is immediately converted into a digital signal (bit-stream) through an A/D converter for digital image processing. In the proposed concept, an additional operation is introduced: analog image pre-processing. It is employed between sensor data acquisition and the A/D-converter (see Fig. 1(b)) and works in the analog domain. Therefore, the image quality can be enhanced in the analog domain at the front end of an image sensor, because the quantization noise from the A/D converter has not yet been accounted for.

For the proposed image pre-processing, the analog input signal is transferred with discrete time, and this transfer occurs sequentially, pixel-by-pixel and line-by-line. Processing with all pixel values in a large image is impossible for analog processing, because both the access latency and the complexity of temporarily storing the pixel values are very high. Therefore, an image restoration method with spatial filters, which only calculates a small part of an image, is used for this study.

A spatial, non-recursive filter is a well-known image restoration method in the research area of digital image processing [6]. These spatial filters are mainly used as either a low-pass to reduce the random image noise or a high-pass to detect edges. During image signal acquisition, the noise is caused by poor illumination, high temperature, and transmission on the image sensor [8]. In general, this image noise could be regarded as random image noise. Non-random noise, e.g. fixed pattern noise, requires additional processing to calibrate. For this reason, it is not discussed in this paper. Two well-known spatial filters [6] support our study: the average filter and the binomial filter.

To compute these spatial filters in a small size, only local image information with the smallest symmetrical filter size of $3 \times 3$ pixels is required. Fig. 2(a) shows linear spatial filtering using a $3 \times 3$ filter: An image patch includes $3 \times 3$ pixels. The value of the center pixel is $f(x,y)$, whereby $x$ and $y$ are the row and column addresses with a sensor size of $m \times n$ pixels. The value of the center pixel for the $3 \times 3$ filter is $w(s,t)$, whereby $s$ and $t$ are the row and column addresses from $-1$, $0$ to $1$. After their convolution, the new value $f'(x,y)$ is calculated for the center pixel $(x,y)$:

$$f'(x,y) = f(x,y) * w(x,y) = \sum_{s=-1}^{1} \sum_{t=-1}^{1} f(x-s, y-t) w(s,t) \quad (1)$$

To compute $f'(x,y)$, a multiplier to multiply $f(x,y)$ with

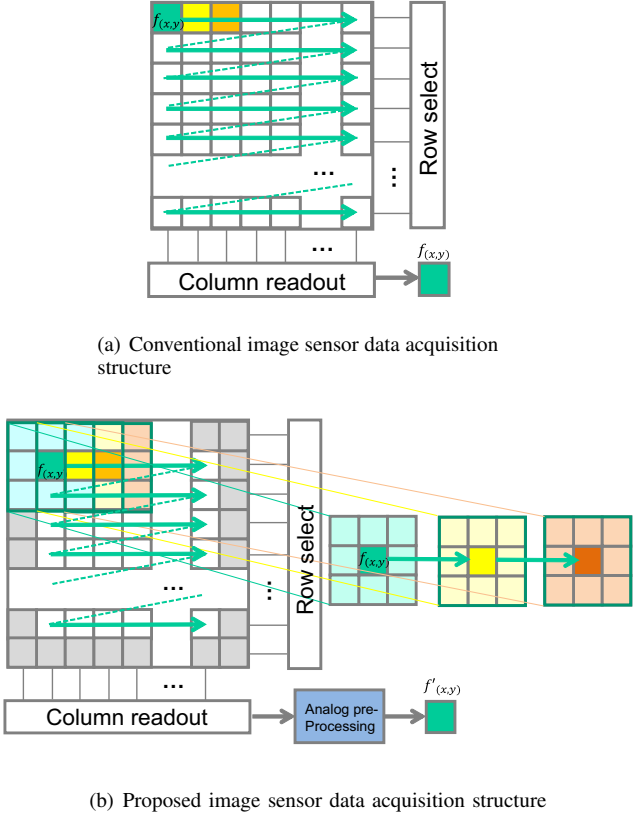

(a) Conventional image sensor data acquisition structure

(b) Proposed image sensor data acquisition structure

Fig. 3. Comparison of image sensor data acquisition structures

$w(x, y)$ and an adder to sum all terms are required. On the analog circuit (see Fig. 2(b)), a combination of a summing amplifier, which sums several (weighted) voltages and yields a negative value, and an inverting amplifier can be performed as follows:

$$V_{out} = \frac{R_{f2}}{R_i}(\frac{R_{f1}}{R_1}V_1 + \frac{R_{f1}}{R_2}V_2 + \ldots + \frac{R_{f1}}{R_n}V_n) \quad (2)$$

The input voltages $V_1$ to $V_n$ ($n = 9$) represent all pixel values $f(x, y)$, $\frac{R_{f1}}{R_1}$ to $\frac{R_{f1}}{R_n}$ represent the filter $w(x, y)$ and the output voltage $V_{out}$ is the required solution to $f'(x, y)$. Furthermore, the output value is reset to be a positive value with $\frac{R_{f2}}{R_i} = 1$. Control signals are used to switch between several pre-defined reference resistors to assign the different values to $R_1$ to $R_n$. Based on the operating principle behind a Gilbert cell [9], Han and Sánchez-Sinencio have provided a survey of CMOS multipliers for a real-time analog multiplication of input signals [10]. Moreover, various CMOS analog arithmetic circuits (addition, subtraction, inversion and multiplication) are employed in [11]. Accordingly, an implementation of a linear spatial filter in a small size on an analog CMOS circuit is feasible.

### III. 2D DATA ACQUISITION METHOD FOR LOW DELAY ANALOG IMAGE PRE-PROCESSING

Based on image pre-processing in the analog domain mentioned in Section II, the pixel values of three adjacent rows and three adjacent columns of a sensor array in 2D need to be accessed in parallel. However, a conventional sensor data acquisition method is sequential, causing immense data input delay for pre-processing, since the pixel value is read out pixel-by-pixel and line-by-line, and the size of the sensor array ($m \times n$) of an image sensor in a current digital camera is far greater than the proposed filter size ($3 \times 3$). More specifically, an image sensor has a "row select and a column readout" structure (see Fig. 3(a)). The signal readout order is left-to-right, top-to-bottom and sequentially pixel-by-pixel. If the readout time of a pixel $(x, y)$ is $\tau$, then the readout time from pixel $(x, y)$ and $(x, y+1)$ in the same row is $2\tau$. Unfortunately, the readout time from the pixel $(x, y)$ and $(x+1, y)$ in the same column takes $(n+1)\tau$. Accordingly, a complete readout of $3 \times 3$ pixels takes $(2n+3)\tau$, and $m, n \gg 3$.

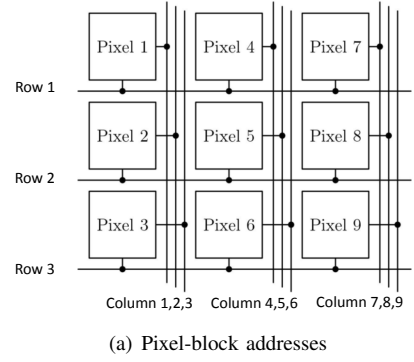

(a) Pixel-block addresses

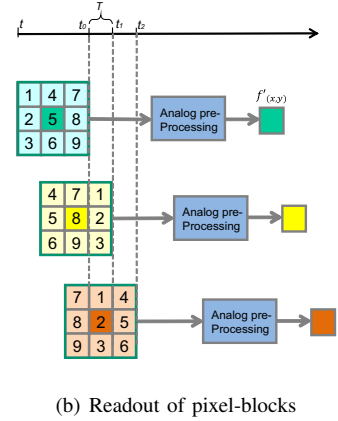

(b) Readout of pixel-blocks

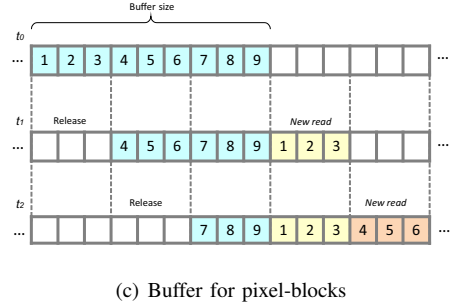

(c) Buffer for pixel-blocks

Fig. 4. Image sensor data addressing, readout and buffering

To reduce this input delay, a novel readout method for 2D

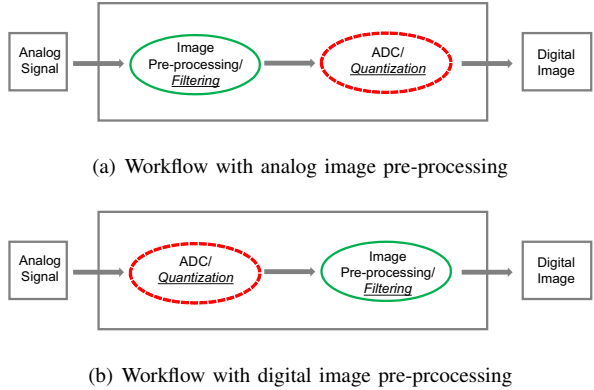

(a) Workflow with analog image pre-processing

(b) Workflow with digital image pre-prcocessing

Fig. 5. Analog-to-digital data transfer workflows with image pre-precessing

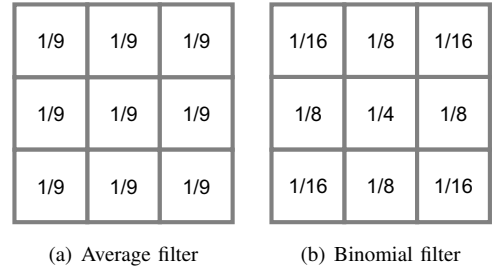

(a) Average filter

(b) Binomial filter

Fig. 6. Spatial filters for reducing random image noise with filter size of $3x3$ pixels

data acquisition for the proposed analog image pre-processing is proposed as follows:

The readout sequence focuses on each $3 \times 3$ pixel cluster. As shown in Fig. 3(b), the image sensor is abstracted to three different levels: pixel, pixel block and image, a pixel being the smallest unit of an image sensor. A pixel block is the same size as the filter, and contains $3 \times 3$ pixels in a cluster, which is the input for the analog image pre-processing. The center of the first pixel block is addressed as $(x, y)$ and the pixel value is $f(x, y)$. An image is comprised of the whole sensor area of $m \times n$ pixels. In the conventional image sensor data acquisition structure, the data acquisition is described as a sequential movement of a pixel one pixel at a time. Our proposed acquisition method, however, is described as a sequential movement of a pixel block with the same step length (one pixel) in the area of the image. The center of a pixel block is the same pixel $(x, y)$. To keep the pixel block's size consistent, the boundary pixels must not be designated as center pixels. In this research, the boundary problems are avoided.

Three parallel lines from each column can be designed in the pixel array (see Fig. 4(a)). With this method, nine pixels in a pixel-block can be read out at the same time, and the readout delay of a pixel-block is reduced to $\tau$. In comparison with the delay $(2n+3)\tau$ ($n \gg 3$), the proposed data acquisition method results in lower input delay from the pre-processing. For each movement of the pixel-block, only three pixels in a column are to be read anew (see Fig. 4(b)). Therefore, a circular buffer [12] [13] with a size of nine values can read three pixels of the next column and releases three pixels from the last column with each pass (see Fig. 4(c)). These buffered values are multiplied with the values of the selected spatial filter and summarized at the next stage: analog pre-processing (see Fig. 3(b)), which was discussed in Section II. The newly calculated analog pixel value $f'(x, y)$ is converted into the digital domain for further digital processing.

## IV. COMPARISON: ANALOG VS. DIGITAL IMAGE PRE-PROCESSING

With the image sensor data acquisition structure proposed above, it is now possible for image pre-processing, such as a 2D-filter, to be applied in the analog domain with low delay data input. In order to compare the image processing performance in the analog and digital domain, two analog-to-digital data transfer workflows are introduced, which contain an identical image signal pre-processing method, e.g., a linear spatial filter, for reducing image random noise, and an identical A/D converter for quantizing the image data into an identical grayscale of digital images. Fig. 5(a) shows the analog image pre-processing and Fig. 5(a) the digital image pre-processing. Their difference is the order of processing and the A/D converter.

To assess the image quality of the output signals through both workflows, 10 test images (see Fig. 8) [14] were used to simulate the ideal system under ideal conditions in MATLAB. For this simulation, the 16-bit test images are regarded as the analog signal in the input stage, and the 8-bit images as the digital signal in the output stage, assuming that a 16-to-8-bit quantization function is an error-free A/D converter. As stated in Section II, two well-known spatial filters, an average filter (see Fig. 6(a)) and a binomial filter (see Fig. 6(b)), are introduced in the assessment as examples. The combination of this pre-processing and A/D converter (in the black square frame of both workflows in Fig. 5) is regarded as the cause for system noise.

Tables I and II show the evaluation results with two image quality measurements: Peak signal-to-noise ratio (PSNR) and structural similarity (SSIM) [15]. With both the average and binomial filter, we determined that simulated analog image pre-processing always resulted in higher quality (higher PSNR and SSIM), producing lower system noise. On average, the output images of the workflow with analog image pre-processing have a higher image quality of $0.88 \ dB$ PSNR with the average filter and $0.42 \ dB$ with the binomial filter in comparison to the workflow with digital pre-processing. Similarly, on average, analog pre-processing improved SSIM (values between $-1$ and $1$) by $0.12$ more than digital pre-processing with an average filter, and $0.03$ with a binomial filter in the simulated workflows. These results reveal that the proposed analog de-noising process is more effective than

19

digital de-noising process.

TABLE I
OBJECTIVE MEASUREMENT RESULTS - PSNR

|  | Average filter | | Binomial filter | |
| --- | --- | --- | --- | --- |
|  | Digital | **Analog** | Digital | **Analog** |
| a. Academy | 34.55 dB | **34.78 dB** | 36.90 dB | **37.07 dB** |
| b. Arri | 33.97 dB | **34.33 dB** | 36.50 dB | **36.69 dB** |
| c. Church | 35.08 dB | **35.58 dB** | 37.71 dB | **37.96 dB** |
| d. Color | 37.71 dB | **40.11 dB** | 41.17 dB | **42.15 dB** |
| e. Face | 38.81 dB | **41.37 dB** | 42.37 dB | **43.32 dB** |
| f. Tree | 30.74 dB | **30.86 dB** | 33.04 dB | **33.14 dB** |
| g. Night | 39.76 dB | **41.09 dB** | 42.24 dB | **43.15 dB** |
| h. Pool | 34.95 dB | **35.26 dB** | 37.36 dB | **37.58 dB** |
| i. Chart | 28.62 dB | **29.12 dB** | 31.43 dB | **31.56 dB** |
| j. Lake | 34.87 dB | **35.37 dB** | 37.36 dB | **37.68 dB** |
| Average gain | - | **0.88 dB** | - | **0.42 dB** |

TABLE II
OBJECTIVE MEASUREMENT RESULTS - SSIM

|  | Average filter | | Binomial filter | |
| --- | --- | --- | --- | --- |
|  | Digital | **Analog** | Digital | **Analog** |
| a. Academy | 0.995 | **0.999** | 0.998 | **0.999** |
| b. Arri | 0.914 | **0.992** | 0.974 | **0.995** |
| c. Church | 0.926 | **0.998** | 0.980 | **0.998** |
| d. Color | 0.716 | **0.970** | 0.902 | **0.979** |
| e. Face | 0.516 | **0.985** | 0.927 | **0.990** |
| f. Tree | 0.978 | **0.997** | 0.992 | **0.998** |
| g. Night | 0.875 | **0.991** | 0.954 | **0.994** |
| h. Pool | 0.969 | **0.998** | 0.991 | **0.999** |
| i. Chart | 0.835 | **0.897** | 0.878 | **0.906** |
| j. Lake | 0.899 | **0.996** | 0.968 | **0.998** |
| Average gain | - | **0.12** | - | **0.03** |

## V. COMPARISON: AVERAGE VS. BINOMIAL FILTER

The proposed concept for a CMOS image sensor incorporates a new data acquisition method with a 2D readout structure. Accordingly, it enables low delay image pre-processing in the analog domain. Furthermore, the results in Section IV show that the output signal offers lower system noise under ideal conditions, so that the image quality can be undoubtedly enhanced by analog image pre-processing compared to digital image pre-processing through linear spatial filters in the analog domain.

As mentioned in Section II, the average filter and the binomial filter were used to remove random image noise. The average filter smooths the image more than the binomial filter because of its uniform value $w(s,t)$ on the filter. However the binomial filter retains more detail than the average filter. To investigate how a filter responds to image noise, a White Gaussian Noise is added to the test images in Fig. 8 for simulating an input image with random noise. As illustrated in Fig. 5(a), the noisy image is processed with the same analog pre-processing but with different filters. Depending on the variance $\sigma^2$ of the added image noise, intersections (colored points in Fig. 7) in the PSNR value always occur in the output stage of the filter. For the test images, if the noise variance $\sigma^2 > 10^{-2}$, the average filter is more effective; if $\sigma^2 < 10^{-3}$, the binomial filter is more effective and provides higher PSNR values; for the range of $10^{-3}$ to $10^{-2}$, the filter effectiveness depends on the input image. Therefore, according to the raw image noise level after sensor data acquisition in the analog domain, an appropriate filter can be determined and designed as an analog image pre-processing unit on the circuit with the proposed concept for a CMOS image sensor.

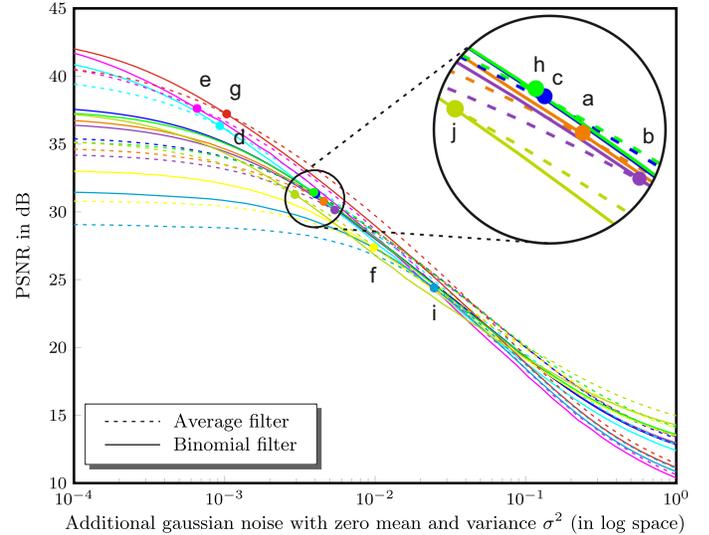

Fig. 7. The average filter vs. binomial filter for analog pre-processing (The compared filters on a same test image are illustrated with same line color but different line style. The color dots are the interception points on each line pair. Letters point to the 10 test images).

## VI. CONCLUSION AND FURTHER RESEARCH

In this paper, a novel concept for a CMOS image sensor was presented. Analog image pre-processing, which could be a spatial filter such as an average or binomial filter, reduces the random image noise in the analog domain under the limit of CMOS analog circuits. To speed up pre-processing, a new sensor signal acquisition method provides a 2D-readout sequence for small pixel blocks with low latency. As indicated via a comparison of the results, analog image pre-processing provides better image quality than digital pre-processing with ideal signal and system conditions. Furthermore, two linear spatial filters for reducing random image noise can be applied to various analog circuit designs depending on the image sensor noise in the analog domain.

In conclusion, the concept proposed for a CMOS image sensor has the potential to improve raw image quality on-chip without greathy reducing processing speed, reducing the pixel's fill factor or demanding external access to other hardware such as FPGAs.

Further research will investigate the CMOS-integrated circuit design of the proposed image sensor data acquisition method and the analog image pre-processing unit in the analog domain. The analog circuit noise and inaccuracies, especially in arithmetical circuits, will be taken into consideration. A self-adaptive filter could also be developed as an enhancement to this study at the next step. Moreover, future studies will encompass applying additional image signal pre-processing in

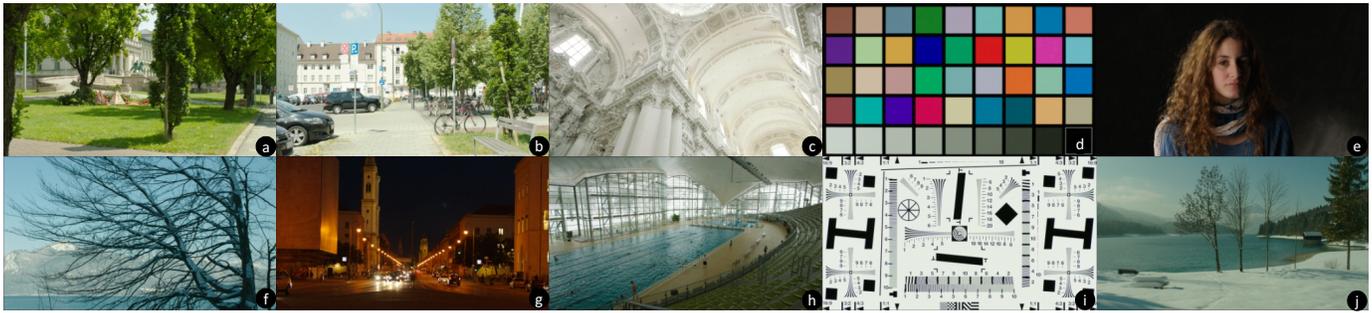

Fig. 8. Test Images: (a) Academy. (b) Arri. (c) Church. (d) Color. (e) Face. (f) Tree. (g) Night. (h) Pool. (i) Chart. (j) Lake. (*Source* [14] *: Freely available raw sensor data from ARRI-Arnold and Richter Cine Technik GmbH*)

the analog domain, e.g., tone mapping operators and color reconstruction.

The proposed concept of image sensors can be beneficial for image sensor research with CMOS circuit design in the area of efficient, ultra-HD, real-time and small-scale chips for digital still or video cameras.

ACKNOWLEDGMENT

The authors would like to thank the Deutsche Forschungsgemeinschaft (DFG) for funding this project (GRK 1773 "Heterogeneous Image Systems").